\documentclass[sn-mathphys]{sn-jnl}

\usepackage{physics}

\usepackage[utf8]{inputenc}
\usepackage{tabularray}
\usepackage{longtable}

\begin{document}

\title[Article Title]{Topological superconductors from a materials perspective}

\author*[1,2]{\fnm{Manasi} \sur{Mandal}}\email{manasim@mit.edu}
\equalcont{These authors contributed equally to this work.}
\author[1,3]{\fnm{Nathan C.} \sur{Drucker}}
\equalcont{These authors contributed equally to this work.}
\author[4]{\fnm{Phum} \sur{Siriviboon}}
\equalcont{These authors contributed equally to this work.}
\author[1,2]{\fnm{Thanh} \sur{Nguyen}} 
\author[1,2]{\fnm{Tiya} \sur{Boonkird}} 
\author[1,2]{\fnm{Tej Nath} \sur{Lamichhane}} 
\author[1,5]{\fnm{Ryotaro} \sur{Okabe}} 
\author[1,6]{\fnm{Abhijatmedhi} \sur{Chotrattanapituk}} 
\author*[1,2]{\fnm{Mingda} \sur{Li}}\email{mingda@mit.edu}

\affil[1]{\orgdiv{Quantum Measurement Group}, \orgname{MIT}, \orgaddress{\city{Cambridge},  \state{MA}, \postcode{02139}, \country{USA}}}
\affil[2]{\orgdiv{Department of Nuclear Science and Engineering}, \orgname{MIT}, \orgaddress{\city{Cambridge}, \state{MA}, \postcode{02139}, \country{USA}}}
\affil[3]{\orgdiv{School of Engineering and Applied Sciences}, \orgname{Harvard University}, \orgaddress{\city{Cambridge},  \state{MA}, \postcode{02138}, \country{USA}}}
\affil[4]{\orgdiv{Department of Physics}, \orgname{MIT}, \orgaddress{\city{Cambridge},  \state{MA}, \postcode{02139}, \country{USA}}}
\affil[5]{\orgdiv{Department of Chemistry}, \orgname{MIT}, \orgaddress{\city{Cambridge},  \state{MA}, \postcode{02139}, \country{USA}}}
\affil[6]{\orgdiv{Department of Electrical Engineering and Computer Science}, \orgname{MIT}, \orgaddress{\city{Cambridge}, \state{MA}, \postcode{02139}, \country{USA}}}

\abstract{Topological superconductors (TSCs) have garnered significant research and industry attention in the past two decades. By hosting  Majorana bound states which can be used as qubits that are robust against local perturbations, TSCs offer a promising platform toward (non-universal) topological quantum computation. However, there has been a scarcity of TSC candidates, and the experimental signatures that identify a TSC are often elusive. In this perspective, after a short review of the TSC basics and theories, we provide an overview of the TSC materials candidates, including natural compounds and synthetic material systems. We further introduce various experimental techniques to probe TSC, focusing on how a system is identified as a TSC candidate, and why a conclusive answer is often challenging to draw. We conclude by calling for new experimental signatures and stronger computational support to accelerate the search for new TSC candidates.}


\maketitle

\tableofcontents

\section{Introduction}\label{sec1}
The field of topological materials has garnered significant research attention over the past decade \cite{Hasan_Kane_RMP2010,Qi_Zhang_RMP2011,Alicea_2012,Manna_NRM_2018,Armitage_RMP2018}. Setting aside the kaleidoscope of fundamental new phenomena emerging from various topological phases, many promising applications have been demonstrated at the lab scale. This includes electronic states with no energy dissipation such as quantum spin Hall effect \cite{Bernevig_QSHE_Science2006,Konig_Science_2007} and quantum anomalous Hall effect \cite{Chang_QAHE_2013,Chang_2016}, current induced switching for spintronic applications \cite{Luqiao_PRL_2017,Luqiao_APL_2021}, topological dipoles for next-generation photovoltaic and photodetectors \cite{LiangFu_PRL_2015,QiongMa2019},  high-efficiency thermoelectrics \cite{Skinner_SciAdv_2018,Han_NC_2020}, catalysis for water splitting and other energy conversion and storage processes \cite{Luo_NatRevPhys_2022}, among others. The recognition of topological materials, or topological phases of matter (since many phases have not been materialized yet) with the 2016 Nobel Prize in Physics was a major milestone \cite{Haldane_RMP_2017,Kosterlitz_RMP_2017}. 

\begin{figure}[ht!]
    \centering
    \includegraphics[width=1.0\columnwidth]{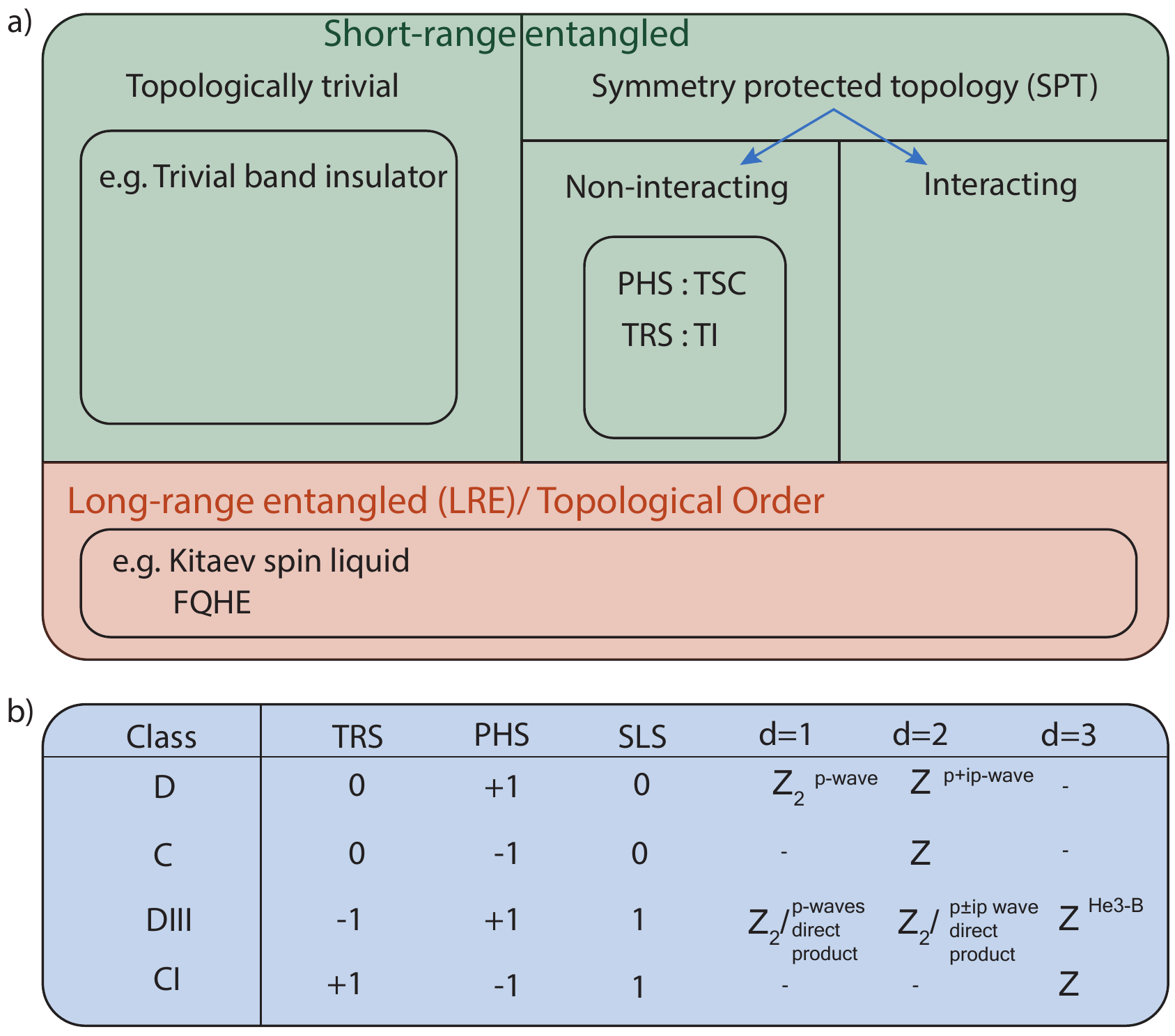}
    \caption{\textbf{Classifications of gapped topological phases of matter and the TSC topological classes.} (\textbf{a}) The TSC family in a zoo of topological materials families. At a mean-field level, TSC can be considered as one type of non-interacting SPT phase. (\textbf{b}) Four subclasses of BdG family topological materials with inherent particle-hole symmetry. The most common TSC is the class D, and a related DIII is a TRS-preserved version which can be considered as direct product of two copies class-D TSCs with opposite chirality. Subfigure (\textbf{b}) adapted from Ref. \cite{Schnyder_10fold_PRB2008}. TRS: Time-reversal symmetry. PHS: Particle-hole symmetry. TI: Topological insulator. FQHE: Fractional quantum Hall effect. SLS: Sub-lattice symmetry. }
    \label{Fig1_TSC_Class}
\end{figure}

Topological superconductors (TSCs) are one class of topological materials and can host Majorana bound states (MBSs, if focusing on the spatial feature), aka Majorana zero modes (MZMs, if focusing on the energy), which can be used as qubits for topological quantum computation. We provide a step-by-step guide to locate the TSC family within the rich families of topological materials (Fig. \ref{Fig1_TSC_Class}a). First, the (gapped) topological phases can be classified by symmetry constraint. Systems hosting topology without any symmetry constraint are intrinsic topological orders with long-range entanglement, while systems with symmetry constraint are short-range entangled \cite{Wen_RMP_2015}. Second, the short-range-entangled states can be classified as either topologically trivial, or topology protected by symmetry, termed symmetry-protected topological (SPT) phases. Third, within the SPT phases, it can further be divided as non-interacting or interacting depending on the strength of electron correlations. The non-interacting SPT phases include the popular topological insulators (TIs) protected by time-reversal symmetry (TRS), topological crystalline insulators (TCIs) protected by crystal symmetry, among others. For interacting SPT phases, the effect of strong electron correlation enriches the phase diagrams \cite{Wang_Senthil_2013,Wang2014a_Senthil,Senthil2015}. It is noteworthy to mention that although Weyl semimetals are non-interacting fermions with topological nodes, they are gapless topological phases that do not belong to the SPT phases discussed here. The gap between the ground state and the excited state allows well-defined collective excitations of ground states and is essential for the robustness against small perturbations. Finally, a superconductor certainly contains strong electron-electron interaction, yet at a mean-field theory level where the electron interaction is approximated as an effective potential (to be discussed in Section \ref{sec2}), a TSC can be classified within the non-interacting SPT family, similar to a TI. Fig. \ref{Fig1_TSC_Class}a provides such a hierarchical structure of overview. Moreover, there are at least four sub-categories belonging to the superconductor and superfluid ``Bogoliubov de Gennes (BdG)" symmetry family (Fig. \ref{Fig1_TSC_Class}b) \cite{Schnyder_10fold_PRB2008}, though it is a misnomer to only consider BdG family as superconductor since some TI family can support fully gapped quasiparticles which can also describe a superconductor (the vice versa is true, that BdG family can also represent non-conventional TI). Within the BdG family, class D breaking TRS can host the 1D $p$-wave TSC and 2D $p+ip$ TSC, and is often considered synonymous with TSC, although many other TSC families exist (such as class DIII which contains two copies of chiral $p$-wave SC with opposite chirality in 1D, direct product of $p \pm ip$ wave SC in 2D, and He3-B phase in 3D). We refer to References \cite{Schnyder_10fold_PRB2008,Ryu_2010_10foldreview} to clarify the complexities.

\begin{figure}[ht!]
    \centering
    \includegraphics[width=1.0\columnwidth]{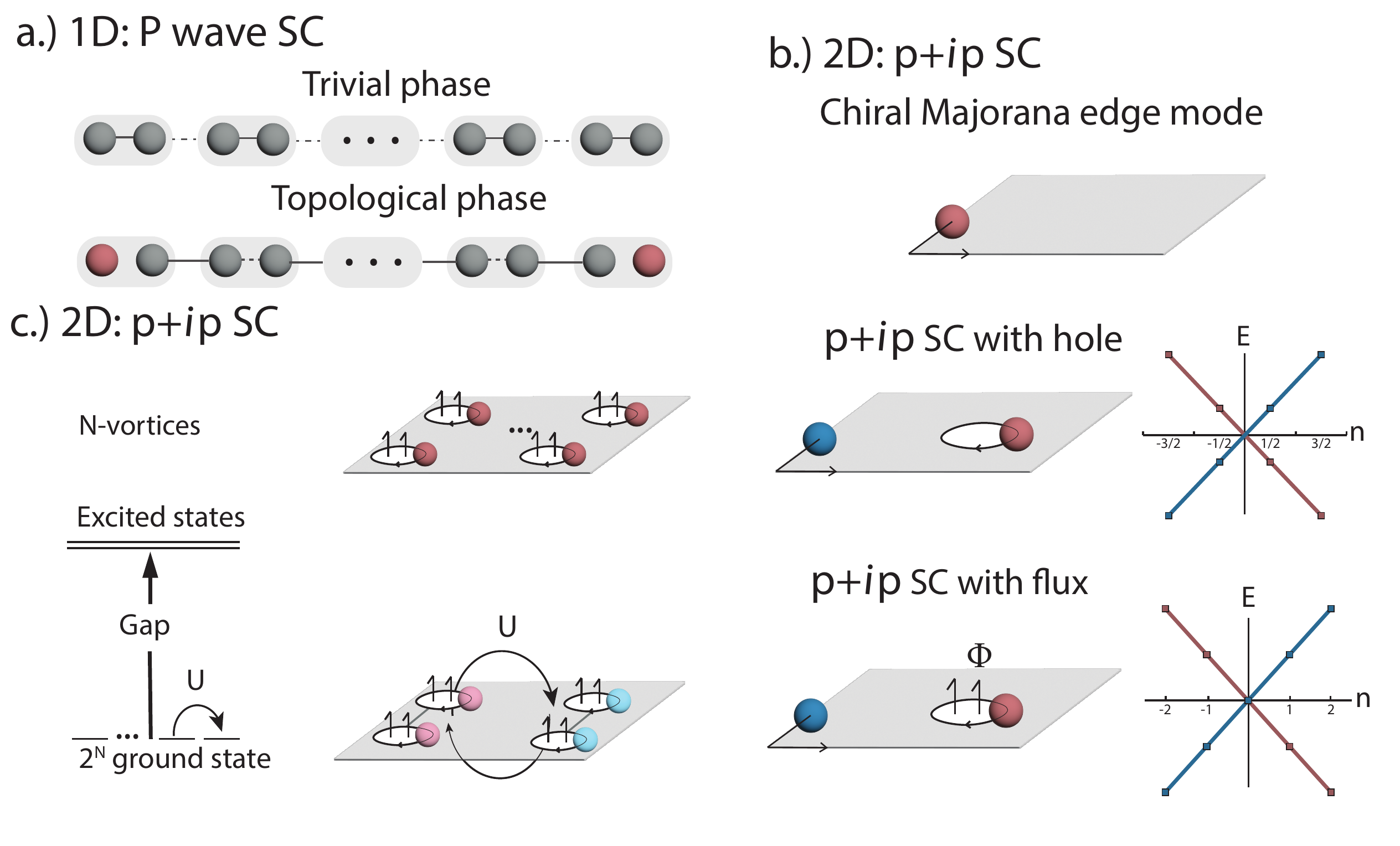}
    \caption{\textbf{Schematic illustration of TSCs and Majorana-based topological quantum computing.} (\textbf{a}) 1D topological superconductor (Kitaev chain), where each conventional fermion is the combination of two Majorana fermions. When ``intra-site" pairing between the two Majorana fermions is stronger than the ``inter-site" pairing (upper), a topologically-trivial SC is obtained. When inter-site interaction is stronger (lower), a 1D TSC is obtained, with two unpaired Majorana fermions (red spheres) with zero energy at two ends. (\textbf{b}) 2D $p+ip$ superconductor. (top) Just like 1D TSC can have 0D boundary modes at two ends, the 2D TSC has 1D chiral Majorana edge modes. (middle) If we pierce one hole to create a region without superconductivity, half-integer excitation spectra are created. (bottom) If we add one magnetic flux quantum $\Phi$ to the hole to create a superconducting vortex, the energy spectra become integers and a Majorana zero mode is generated. (\textbf{c}) Scheme for topological quantum computation. With 2$N$ superconducting vortices, the ground states will have a 2$^{N}$ degeneracy. The unitary transform of U, which can be used as a quantum gate, can be realized by exchanging different pairs of Majorana zero modes within the ground states.}
    \label{Fig2_Schematic}
\end{figure}

 Majorana fermions can emerge in a TSC (but not in a conventional metal) because a Majorana fermion is its own antiparticle. In conventional metals and insulators, a quasiparticle (such as an electron or a hole) carries an electrical charge, with their antiparticle having an opposite charge. Therefore, Majorana fermions are unlikely to emerge as quasiparticles in metals and insulators. A superconductor is therefore a better platform to search for Majorana fermions because of particle-hole symmetry. However, conventional $s$-wave superconductors are also unlikely to host Majorana fermions since the quasiparticle, although formed as a superposition of an electron and a hole, contains opposite spins between the electron and hole, and thus cannot be its own antiparticle: an anti-quasiparticle will have electron and hole with opposite spin states. This leaves the chiral $p$-wave, odd-pairing superconductors (in 1D) and $p + ip$ pairing superconductors (in 2D) as natural choices to search for Majorana fermions. Even so, the 1D $p$-wave and 2D $p + ip$ TSC, which can host Majorana fermions, are only one sub-class of the TSC, and vice versa, there are other classes that can hold Majorana fermions (e.g. DIII class in 3D can host 2D surface Majorana fermion modes). Moreover, it is a misnomer to state that quantum computation is based on Majorana fermions, since Majorana fermions are still fermions satisfying conventional fermionic statistics. To achieve quantum computing, a Majorana fermion normally needs to be bound to a defect (hence called MBS) with zero energy (hence also called MZM), which can contain nontrivial non-Abelian statistics. This review uses MBS and MZM interchangeably. To clarify, a Marjoana fermion binding a vortex core in a 2D $p+ip$ superconductor is not the only way to form the zero modes; there are other approaches to form localized zero modes when bound to topological defects \cite{Ryu_2010_10foldreview}. Here, an MZM is a type of anyon termed Ising anyon, which contain non-Abelian statistics but by itself is not sufficient to carry out universal quantum computations \cite{Alicea_2012,Nayak_RMP_2008}. Fig. \ref{Fig2_Schematic} explains the details of how MBS is formed in a TSC and how quantum computing is achieved.

\section{Theory}\label{sec2}

\subsection{Superconductivity Crash Course}
\begin{figure} 
    \centering
    \includegraphics[width=0.5\linewidth]{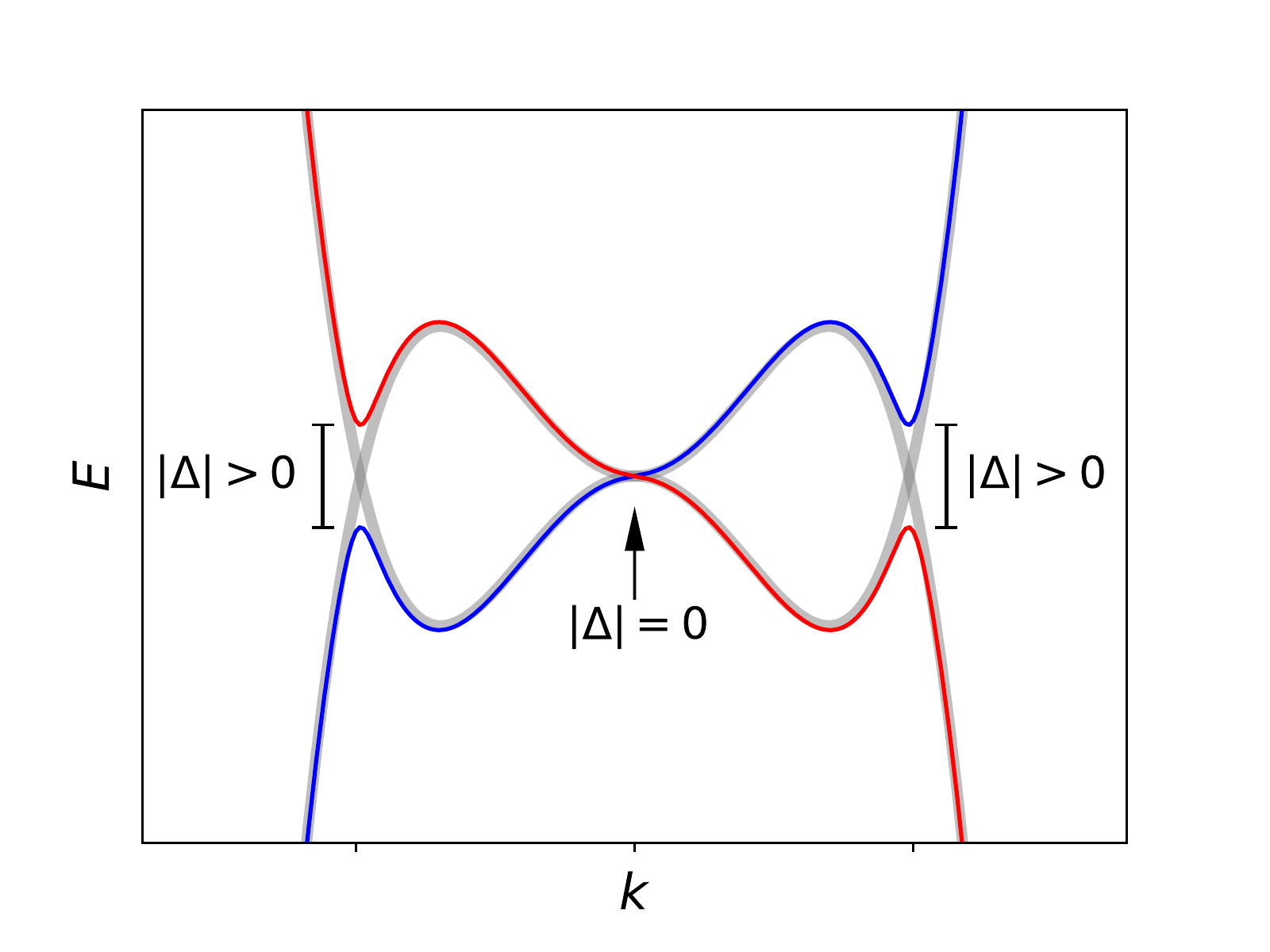}
    \caption{\textbf{An example of the superconductor bandstructure.} The gray lines indicate the energy band $\epsilon(k), -\epsilon(-k)$ while the red and blue lines are the bandstructure after tuning on the interaction potential. Note that since $\Delta$ is momentum-dependent, the perturbation of band crossing can result in  either the nodal or gap structure.}
    \label{fig:TSC-band}
\end{figure}

A simple model for a superconductor can be described by single-band Hamiltonian with a two-body mean field interaction \cite{Bernevig+2013}
\begin{align} \label{eq:1}
    H = \sum_{k, \alpha, \beta} \epsilon_{\alpha, \beta}(k) c^{\dagger}_{k\alpha} c_{k\beta}  + \sum_{k, \alpha, \beta} \left ( \Delta_{\alpha, \beta}(k) c^\dagger_{-k\alpha} c^\dagger_{k \beta}  +h.c. \right )
\end{align}
where $c^\dagger_{k\alpha}$, $c_{k\alpha}$ is creation/annihilation operator of particle at crystal momentum $k$ and spin $\alpha$, and $\epsilon_{\alpha, \beta}(k)$ is measured from the chemical potential. The mean field potential $\Delta_{\alpha\beta}(k)$ acts as an attractive potential that binds electrons together into the Cooper pair state and also serves as the order parameter. Due to the fermionic statistics of electrons, the potential must obey the following constraint
\begin{align} \label{eq:2}
    \Delta_{\alpha, \beta} (k) = - \Delta_{\beta, \alpha} (-k).
\end{align}
In other words, the pairing potential must be anti-symmetric in momentum space (triplet pairing) or spin space (singlet pairing). To understand the bandstructure picture of the superconductor, it is helpful to introduce the Bogoliubov-de-Gennes (BdG) transformation of basis $\begin{pmatrix} c^\dagger_{k}&  c_{-k}\end{pmatrix}$ where we can rewrite the Hamiltonian as

\begin{align} \label{eq:3}
    H_{BdG} = \begin{pmatrix}
        \epsilon(k) & \Delta(k) \\
        \Delta^\dagger(k) & - \epsilon(-k)
    \end{pmatrix}
\end{align}
and we can see that $\epsilon(k)$ and $-\epsilon(-k)$ are related by particle-hole transformation. By treating $\Delta(k)$ perturbatively, we can consider the bandstructure as $\epsilon(k), -\epsilon(-k)$ which gap out at the band crossing point $q$ with the energy gap $\pm \lvert \Delta(q)\rvert$ (Fig. \ref{fig:TSC-band}). It is worth noting that the quasiparticle excitations in this system, the BdG quasiparticles, are the superposition of the particle and hole states discussed in the Section above.

\subsection{Majorana Zero Modes}

Here, we consider Kitaev's model \cite{kitaev2001unpaired} to show the emergence of MZMs in the 1D $p$-wave superconductor 
\begin{align} \label{eq:4}
    H = \sum_{j} -t (c^\dagger_j c_{j+1} + h.c.) - \mu c^\dagger_j c_j + \lvert\Delta\rvert (c^\dagger_j c^\dagger_{j+1} + h.c.)
\end{align}
where we note that the superconductor pairing term is momentum-dependent due to its cross-site pairing. In the Bloch basis, the Hamiltonian can be written as
\begin{align} \label{eq:5}
    H_{Bloch}(k) = \begin{pmatrix}
        -2 t \cos(k) - \mu & 2 i \lvert \Delta \rvert \sin(k) \\ 
        - 2 i \lvert \Delta \rvert \sin(k) & 2 t \cos(k) + \mu
    \end{pmatrix}
\end{align}
with energy spectrum $ E= \pm \sqrt{(2t \cos(k) + \mu)^2 + 4 \lvert \Delta \rvert^2 \sin^2(k)}$. We point out that at $\mu = -2t$ and $2t$, the spectrum becomes gapless, allowing the topological transition of the band. Consider the limiting case where $\lvert \mu \rvert \to \infty$, the eigenstate of the system becomes either localized in the lattice site (for $\mu 
 \to -\infty$) or completely empty as vacuum (for $\mu 
 \to +\infty$), i.e., the system is a trivial atomic insulator (Fig. \ref{Fig2_Schematic}a, upper). On the contrary, the scenario where $-2t < \mu <  2t$ is indeed topological (Fig. \ref{Fig2_Schematic}a, lower). Due to the bulk-boundary correspondence, the boundary between such topological state and topologically trivial state would host boundary modes to reconcile the topological number discontinuity  \cite{sato2017topological}. To reveal the features of MBS, we can define the Majorana operator $a_j$ from the fermion operator $c_j$ 
\begin{align} \label{eq:6} 
    c_j &= \frac{1}{2} (a_{2j-1} +  a_{2j})\\
c_j^\dagger &= \frac{1}{2} (a_{2j-1} -  a_{2j}) 
\end{align}

We notice that the self-conjugating property $a_j = a_j^\dagger$ and the anti-commutation relation $\{a_i, a_j\} = 2 \delta_{ij}$ of the Majorana operator are sufficient to preserve the anti-commutation relationship of fermion $\{c^\dagger_i, c_j\} = \delta_{ij}$ and $\{c^\dagger_i, c^\dagger_j\} = \{c_i, c_j\} = 0$. The Hamiltonian can then be written as 

\begin{align} \label{eq:8}
    H = \frac{i}{2} \sum_j - \mu a_{2j-1} a_{2j} + (t + \lvert \Delta \rvert) a_{2j} a_{2j+1} + (-t + \lvert \Delta \rvert) a_{2j-1} a_{2j+2}.
\end{align}

This leads back to the schematics in Fig. \ref{Fig2_Schematic}a where $\mu$ acts as the ``intra-site" interaction, i.e., hopping term between the two Majorana fermions at the same site, and $t + \lvert \Delta \rvert$ and $-t + \lvert \Delta \rvert$ correspond to the hopping across neighboring sites, i.e., ``inter-site" hopping. In the topological trivial state, the Majorana fermion pairs up on the same site and acts as a normal fermion (Fig. \ref{Fig2_Schematic}a, upper figure). In the topological non-trivial case, e.g. $\mu = 0$, $t = \lvert \Delta \rvert > 0$, the Majorana pairing occurs on the neighboring sites of the lattice, resulting in the leftover Majorana states at the two boundaries (red spheres in Fig. \ref{Fig2_Schematic}a, lower figure).

In order to demonstrate computation with such states, we consider a system of superconducting islands with tunable chemical potentials and "valves" connecting the superconductivity \cite{aasen2016milestones}. Note that due to the non-Abelian statistics of the MZMs \cite{ivanov2001non}, the permutation of 2N-MZMs would form a braid group which could be used as a basis for robust quantum computing against local perturbation. The topological robustness can be seen since information can be stored nonlocally (at two ends of the 1D $p$-wave TSC, or two MZMs spatially away from each other in 2D $p+ip$ TSC). Thermal fluctuations at finite temperature could still pose a challenge, since the system could be thermally excited to an excited state (Fig. \ref{Fig2_Schematic}c).

\section{Candidate Materials}
The experimental realization of TSC is currently limited to a relatively small pool of candidate materials. These materials can be broadly categorized into two groups: natural candidates, which may host TSC own their own, and artificial candidates, which require 2D heterostructures or 1D wires with multiple constituent materials.

\subsection{Natural Candidates}\label{subsubsec2}

In the search for TSCs, spin-triplet pairing superconductors which can host MBS have garnered significant attention (DIII family in Fig. \ref{Fig1_TSC_Class}b). A handful of bulk TSC candidates have been reported, which we summarize below. While we have attempted to be as comprehensive as possible, it is important to note that our list may not be fully exhaustive due to our limited knowledge. However, it should still cover a decent portion of known TSC candidates by far.

\textbf{Sr$_{2}$RuO$_{4}$:} 
Superconductivity in strontium ruthenate was discovered in 1994 and continues to draw significant research interest because of its mysterious superconducting pairing symmetry. While the chiral $p$-wave pairing is believed to be realized in the A-phase of this material, favored by several early experiments, such as the muon spin rotation and relaxation ($\mu$SR) \cite{ribak2020} and the Knight shift \cite{ishida1998} measurements taken in 1998. In contrast, later reports by Knight shift \cite{romer2019}, NMR \cite{pustogow2019}, and polarized neutron scattering experiments \cite{steffens2019} conducted in 2019 have raised questions regarding the presence of the “chiral $p$-wave” or even the spin-triplet pairing in this material. The intrinsic Josephson junction \cite{yasui2020}, and the in-plane-magnetic-field stabilized half-quantum vortices \cite {roberts2013} experiments show that the pairing symmetry in Sr$_{2}$RuO$_{4}$ is a triplet with its spin polarization axis aligning in an in-plane direction. A recent report on stress-induced splitting between the onset temperatures of superconductivity and TRS breaking by zero-field-$\mu$SR measurements claimed the qualitative expectations for a chiral order parameter in this mysterious compound \cite {grinenko2021}. Liu \textit{et al.} proposed the following scheme: the order parameter of Sr$_{2}$RuO$_{4}$ is not chiral $p$-wave but instead one of the helical states that do not break TRS overall, yet in which each spin component does break it individually \cite {leggett2021}. More experimental measurements and theoretical work are needed to unambiguously establish the precise pairing symmetry in this TSC candidate.

\textbf{UPt$_{3}$:}
Another leading candidate for bulk TSC is UPt$_{3}$ with multiple superconducting phases \cite{adenwalla1990}. The TRS breaking superconducting state was reported many years ago by $\mu$SR \cite{luke1993} with further confirmation by a Kerr effect study \cite{schemm2014}. The temperature dependence of the upper critical field is reported to exhibit a strong anisotropy \cite{shivaram1986} with a possible strong spin-orbit interaction locking the direction of zero spin projection \cite{choi1991}. These results were contradictory to the NMR results \cite{tou1998} that suggest an equal-spin pairing state with the spin angular momentum directed along the magnetic field, which is possible only in the presence of little or no spin-orbit coupling. Another study by polarized neutron scattering probe predicts odd-parity, spin-triplet superconductivity in UPt$_{3}$ \cite{gannon2012}. The gap symmetry in UPt$_{3}$ was investigated by thermal conductivity tensors where the field-angle-resolved thermal conductivity shows spontaneous twofold symmetry breaking in the gap function for the high-field C-phase, indicating that the pairing symmetry belongs to the $f$-wave category. The theoretically proposed chiral $f$-state is compatible with most of the experimental results reported until now \cite{goswami2015,izawa014}. However, some fundamental issues, such as the existence of tetra-critical points, have not been explained within the scenario of chiral states \cite{izawa014}. First-principles analysis predicted the microscopic superconducting gap structure as an E$_{2u}$ state with in-plane twofold vertical line nodes on small Fermi surfaces and point nodes with linear dispersion on a large Fermi surface \cite{nomoto2016}. A recent report by small-angle neutron scattering evidenced bulk broken TRS in this heavy-fermion superconductor UPt$_{3}$ with anisotropy of the order parameter and current density near the vortex cores \cite{avers2020}.

\textbf{URu$_{2}$Si$_{2}$:}
The pairing mechanism of unconventional superconductivity in the heavy-fermion compound, URu$_{2}$Si$_{2}$, has been a longstanding mystery, despite being intensively studied by several experimental and theoretical groups. Polar Kerr effect \cite{schemm2015}, magnetic torque \cite{li2013}, and $\mu$SR \cite{kawasaki2014} measurements have provided evidence for the bulk TRS broken superconducting state. In addition, the observation of a colossal Nernst signal attributed to the superconducting fluctuations has been reported, where the results were predicted as chiral or Berry-phase fluctuations associated with the broken TRS of the superconducting order parameter \cite{yamashita2015, sumiyoshi2014}. Furthermore, the field-orientation-dependent specific heat measurements and theoretical analyses described the gap symmetry of URu$_{2}$Si$_{2}$ as a chiral $d$-wave-type \cite{kittaka2016}.

\textbf{SrPtAs:}
Hexagonal honeycomb structure SrPtAs superconductor with a spontaneous TRS breaking state was reported based on $\mu$SR experiments, suggesting possible chiral-$d$-wave states \cite{biswas2013}. Recent nuclear magnetic resonance measurements showed multigap superconductivity \cite{matano2014} and a suppressed coherence peak that supports chiral $d$-wave order parameter \cite{bruckner2014}. According to a theoretical study \cite{fischer2014}, SrPtAs is a superconductor with protected Majorana-Weyl nodes in bulk and (Majorana) Fermi arcs on the surface, along with other topological Majorana surface states. However, further experimental evidence is needed to confirm these predictions.

\textbf{UTe$_{2}$:}
The recently discovered heavy-fermion superconductor UTe$_{2}$ is a prime candidate for a topological chiral spin-triplet superconductor. There are several reports on spin-triplet pairing \cite{ran2019,ran2019-2,jiao2020} and the possible chiral state \cite{hayes2021,metz2019}, but the symmetry and nodal structure of the order parameter remain controversial \cite{shishidou2021}. The anisotropy of low-energy quasiparticle excitations indicates that the order parameter has multiple components in a complex chiral form, which provides hints of the topological properties in UTe$_{2}$ \cite{ishihara2022}. More intriguingly, the optical Kerr effect \cite{hayes2021}, and microwave surface impedance measurements \cite{bae2021} suggest a TRS broken superconducting state. A scanning tunneling microscopy (STM) study reveals signatures of chiral in-gap states, suggesting UTe$_{2}$ is a strong candidate for chiral-triplet TSC \cite{jiao2020}.

\textbf{Transition Metal Dichalcogenides:}
The superconductivity of 2M-WS$_{2}$, a transition metal dichalcogenide (TMD), was recently confirmed by transport measurements \cite{2M-WS2} and scanning tunneling microscopy/spectroscopy (STM/STS) investigations. Zero energy peaks in the STS spectra were observed in magnetic vortex cores \cite{yuan2019}, suggesting the possible existence of MZMs. A further angle-resolved photoemission spectroscopy (ARPES) study established the TSC nature of 2M-WS$_{2}$ \cite{li2021}. Chiral superconductivity is reported in another TMD, 4Hb-TaS$_{2}$ \cite{ribak2020} with a strong signature of zero-bias states in vortex cores \cite{nayak2021}. 
Other systems, such as MoTe$_{2}$ \cite{luo2016,li2018,guguchia2017}, and WTe$_{2}$ \cite{kang2015} (parent and doped), are also of significant research interest in exploring possible topological superconductivity. However, the direct observation of topological surface states (TSSs) and the superconducting gap of the TSSs are yet to be explored.

\textbf{LaPt$_{3}$P:}
The weakly correlated pnictide compound LaPt$_{3}$P, with centrosymmetric crystal structure, has been reported TRS broken superconducting state and low-temperature linear behavior in the superfluid density, indicating line nodes in the order parameter. It was predicted that LaPt$_{3}$P is a chiral $d$-wave singlet using symmetry analysis, first-principles bandstructure calculation, and mean-field theory \cite{biswas2021}.

\textbf{Doped Topological Insulators:}
One approach to realizing TSC is through doping bulk topological insulators (TIs). This approach is attractive because it has the potential to realize the coexistence between fully gapped, bulk superconductivity and topological surface states. In addition, the strong spin-orbit coupling (SOC) of topological materials may lead to unconventional pairing mechanisms \cite{fu2010}. Early experimental efforts on this front focused on electrochemically intercalating Cu into Bi$_{2}$Se$_{3}$ \cite{hor2010,kriener2011}. Here, superconducting transition temperatures of $T_c \sim 3.8$K were observed for Cu doping between 10\% and 30\%. Moreover, there have been observations of a zero-bias conduction peak on the surface of Cu$_{0.3}$Bi$_{2}$Se$_{3}$ through point contact spectroscopy \cite{sasaki2011}, encouraging the possibility of TSC in this system. However, follow-up STS measurements found no evidence of a zero-bias conductance peak intrinsic to the material and also found evidence for conventional BCS $s$-wave superconducting pairing \cite{Nivy2013}. The search for TSCs in this system has also been hampered by the low percentage of bulk material that actually show superconductivity. To overcome this challenge, there have been efforts to intercalate Sr and Ti instead of Cu, which has resulted in larger bulk superconducting fractions of $\sim$91\% along with transport evidence for topological surface states \cite{liu2015,wang2016}. Along these lines, there have also been investigations on superconductivity resulting from In doping of TI SnTe \cite{erickson2009,sasaki2012,novak2013}. Nevertheless, the pairing mechanism in this class of materials remains under debate.

\textbf{Non-centrosymmetric Superconductors:}
Another class of materials that may host TSC is the non-centrosymmetric superconductors. Due to their broken inversion symmetry, these materials are allowed to have asymmetric spin-orbit coupling (ASOC), which can mix singlet and triplet superconductivity at sufficiently large strength. This is a large family of materials with thorough reviews reported elsewhere \cite{smidman2017}. Here, we highlight some notable examples. A central challenge in this class of materials is unambiguously determining the degree of singlet-triplet pairing. For example, CePt$_{3}$Si \cite{bauer2004} is a heavy fermion superconductor at ambient pressure with Rashba SOC. It exhibits antiferromagnetism below $T_{N}= 2.2$ K, and superconductivity below $T_{c}= 0.7$ K \cite{amato2005}, with thermal transport measurements indicating line nodes in the pairing gap \cite{izawa2005}. However, it is possible that the line node in this system arises due to the coexistence of antiferromagnetic and SC orders \cite{doi:10.1143/JPSJ.75.083704}, highlighting the difficulty in extracting the pairing structure from experiments. More recently, superconductivity has been found in topological, inversion symmetry breaking half Heusler compounds RPdBi (R = Ho, Er, Tm, Lu) \cite{RPdBi} and RPtBi (R = La, Lu, Y) \cite{doi:10.1073/pnas.1810842115, PhysRevB.93.115134}, which are attractive systems because of the chemical and magnetic tunability, as well as the possibility for higher-angular momentum pairing. 

\textbf{Fe-based Superconductors:}
Fe-based superconductors are yet another promising avenue for realizing TSC. Prominently featured within this broad family is the FeSe$_{1-x}$Te$_{x}$ (FTS) system. In this material, topological bands are driven by SOC from Te inclusion.
Unlike other TSC candidates, the unconventional pairing mechanism in FTS \cite{hanaguri2010} is not as hotly debated, and there is evidence for topological surface states coexisting with a hard SC gap \cite{wang2015,zhang2019,hao2014,zhang2018,xu2016}. There have been significant experimental efforts to realize MZM within vortex cores on the surface of FTS \cite{machida2019,kong2019,wang2018,zhu2020} and (Li$_{0.84}$Fe$_{0.16}$)OHFeSe \cite{liu2018}. Nevertheless, it remains a challenge to determine whether the hallmark signature of MZMs, the zero-bias peak (ZBP), is related to trivial states instead of topological states \cite{jeon2017}. More recently, there have been theoretical predictions \cite{zhang2019} and experimental evidence \cite{gray2019} for helical hinge MZMs in FTS arising from the topological nature and $s_{+-}$ pairing mechanism of this material. 

\subsection{Artificial Candidates}

Artificial structures such as engineered heterostructures and nanowire hybrids are currently attracting significant attention as an alternative to intrinsic TSCs. Several recipes came up for the possible realization of TSCs and detecting MZMs. Here, we briefly summarize predicted systems and experimental materials of the past decade, such as\\
1.	1D TSC - Hybrid SC - semiconductor nanowire and magnetic atoms chain/SC\\
2.	2D TSC - Topological insulator (Topological crystalline insulator)/SC heterostructures

\textbf{1D TSC:}
It was proposed that MZMs can emerge at the ends of 1D TSCs. Several studies report possible 1D TSCs, mainly involving hybrid superconductor–semiconductor nanowire devices in the presence of an applied magnetic field along the nanowire axis \cite{das2012,mourik2012,deng2016,nichele2017}. MZMs are expected to arise at each end of the wire in such a system. An aluminum superconductor in proximity to an InAs nanowire having strong SOC and Zeeman splitting is reported to show a distinct zero-bias conductance peak and its splitting in energy, with a small applied magnetic field along the wire \cite{das2012}. Similar predictions were given by another group \cite{deng2016, nichele2017}. Another hybrid structure of indium antimonide nanowires contacted with superconducting niobium-titanium nitride shows bound states at zero bias with the variation of magnetic fields and gate voltages, supporting the hypothesis of Majorana fermions in nanowires coupled to superconductors \cite{mourik2012}.
Nadj-Perge \textit{et al.} created an alternative hybrid system by depositing iron atoms onto the surface of superconducting lead, where enhanced conductance at the ends of these chains at zero energy was observed by STM \cite{nadj-perge2014}. Here proximity-induced superconductivity was expected to be topological due to the odd number of band crossings at the Fermi level. 

\textbf{2D TSC:}
Research on 2D TSCs attracted increased attention after a prediction made by Fu and Kane based on an $s$-wave SC and TI heterostructure. The helical pairing of the Dirac fermions was realized in a Bi$_{2}$Se$_{3}$/Nb heterostructure \cite{flototto2018}. Other heterostructures such as Bi$_{2}$Se$_{3}$/NbSe$_{2}$ and Bi$_{2}$Te$_{3}$/NbSe$_{2}$ were also predicted as possible TSC \cite{sun2016,wang2012,xu2014,xu2015}. However, the proximity-induced pairing potential is very small for the above systems because of the low pairing potential of the $s$-wave superconductor. Later, unconventional superconductors were used instead of $s$-wave superconductors in several systems \cite{eich2016,he2014,chen2018}, such as Bi$_{2}$Te$_{3}$/FeTe \cite{he2014} and Bi$_{2}$Te$_{3}$/FeTe$_{0.55}$Se$_{0.45}$ \cite{chen2018}; the question of whether the induced superconductivity can be regarded as chiral $p$-wave pairing is still under debate. Proximity-induced superconductivity was also investigated in atomically flat lateral and vertical heterostructures of TCI Sn$_{1-x}$Pb$_{x}$Te and superconducting Pb \cite{zhu2021,yang2020}. High-resolution STM measurements make it a promising candidate for TSC. A recent report on the Moiré pattern between a van der Waals superconductor (NbSe$_{2}$) and a monolayer ferromagnet (CrBr$_{3}$) proposed periodic potential modulations arising from the Moiré pattern as a powerful way to overcome the conventional constraints for realizing and controlling topological superconductivity \cite{kezilebieke2022,kezilebieke2020}. Another system, such as gated monolayer WTe$_{2}$, is reported as a higher-order topological superconducting candidate with inversion-protected Majorana corner modes without proximity effect \cite{hsu2020}. Further studies using STM or transport measurements are required to probe Majorana corner modes.

Natural and artificial TSC candidates are summarized in Table \ref{tab1}.

\begin{center}
\begin{longtable}{|p{2.5cm} |p{4.5cm} |p{4.5cm}|}
\caption{A summary of TSC candidates}
\label{tab1}\\
    \hline
    Possible TSC & Features & Drawback\\
    \hline\hline
    \endhead
    Sr$_{2}$RuO$_{4}$ \par \cite{ribak2020,ishida1998,romer2019,pustogow2019,steffens2019,yasui2020,roberts2013,grinenko2021,leggett2021} & 
    Chiral superconductor & 
    Pairing symmetry is still controversial\\
    
    \hline
    
    UPt$_{3}$ \par \cite{adenwalla1990,luke1993,schemm2014,shivaram1986,choi1991,tou1998,gannon2012,goswami2015,izawa014,nomoto2016,avers2020} & 
    1. Time-reversal symmetry-breaking (TRSB) state in bulk \par 
    2. Proposed as chiral $f$-wave SC & 
    Fundamental issues have not been explained within the scenario of chiral states \\
    \hline
    
    URu$_{2}$Si$_{2}$ \par
    \cite{schemm2015,li2013,kawasaki2014,yamashita2015, sumiyoshi2014,kittaka2016}
    &
    1. Bulk TRSB \par
    2. Proposed as chiral $d$-wave SC &
    Pairing mechanism is still a mystery\\

    \hline
    
    SrPtAs \par \cite{biswas2013,matano2014,bruckner2014,fischer2014} &
    1. Multigap superconductor \par
    2. Possible chiral $d$-wave states & 
    Lack of experimental evidence\\

    \hline
    UTe$_{2}$ \par 
    \cite{ran2019,ran2019-2,jiao2020,hayes2021,metz2019,shishidou2021,ishihara2022,bae2021} &
    1. TRSB \par
    2. Complex chiral form in order parameter \par
    3. Prime candidate for chiral spin-triplet superconductor 
    & Symmetry and nodal structure of the order parameter remains controversial\\
    
    \hline
    
    2M-WS$_{2}$ \par
    \cite{2M-WS2,yuan2019,li2021}  
    
    & 
    1. Zero energy peaks in the STS spectra in magnetic vortex cores \par
    2. Topological surface states acquired nodeless superconducting gap &  No direct evidence   \\
    \hline
    
    4Hb-TaS$_{2}$ \par  \cite{ribak2020,nayak2021}
    & Strong signature zero-bias states in vortex cores &   Requires further investigation  \\
    \hline

    MoTe$_{2}$ (parent and doped) \par \cite{luo2016,li2018,guguchia2017} & 
    1. Type-II Weyl semimetal \par 
    2. 2-gap $s$-wave symmetry \par
    3. Suggested topologically non-trivial $s_{+-}$ state \par
    4. MoTe$_{2-x}$S$_{x}$ two-band $s$-wave bulk superconductor & 
    No direct observation of topological surface states (TSSs) and the superconducting gap of the TSSs \\

    \hline

    LaPt$_{3}$P \par \cite{biswas2021} & 
    1. TRSB \par
    2. Low-temperature linear behavior in the superfluid density \par
    3. Predicted as chiral $d$-wave SC & No direct evidence\\

    \hline

    Cu$_{x}$Bi$_{2}$Se$_{3}$ \par \cite{hor2010,kriener2011,sasaki2011} & 
    Zero-bias conduction peak by point contact spectroscopy &
    1. Low superconducting  volume fraction \par 
    2. STM measurements found no evidence of a zero-bias conductance peak intrinsic to the material \par
    3. Evidence for conventional BCS $s$-wave superconducting pairing\\

    \hline

    CePt$_{3}$Si \par \cite{bauer2004,amato2005,izawa2005,doi:10.1143/JPSJ.75.083704} 
    & Line nodes in the pairing gap & Line node may arise due to the coexistence of AFM and SC orders\\

    \hline
    
    Fe-based superconductor \par \cite{hanaguri2010, wang2015,zhang2019,hao2014,zhang2018,xu2016,machida2019,kong2019,wang2018,zhu2020,liu2018,jeon2017,gray2019}  & 
    Topological surface states coexisting with a hard superconducting gap & Challenge to determine whether the hallmark signature of MZM’s, the zero-bias peak (ZBP), is related to trivial states instead of topological states \\

    \hline
    
    1D TSC: Al/ InAs or InSb Nanowire, Fe atoms chain/Pb
    \cite{deng2016, nichele2017,mourik2012,nadj-perge2014}
    
    & Distinct zero-bias conductance peak & Such hybrid systems do not provide definite proof of a Majorana state\\

    \hline
    
    Bi$_2$Se$_3$/Nb, Bi$_{2}$Se$_{3}$/NbSe$_{2}$ and Bi$_{2}$Te$_{3}$/NbSe$_{2}$
    heterostructure \par
    \cite{flototto2018,sun2016,wang2012,xu2014,xu2015}
      
    & Helical pairing of the Dirac Fermions & Proximity-induced pairing potential is very small \\

    \hline
    
    Bi$_2$Te$_3$/\par FeTe$_{0.55}$Se$_{0.45}$ 
    \cite{eich2016,he2014,chen2018}
    & Two-fold superconductivity & Pairing symmetry is under debate\\

    \hline
    
\end{longtable}
\end{center}

\section{Experimental Signatures}\label{sec8}

\subsection{Tunneling Spectroscopy}
\begin{figure}[ht!]
    \centering
    \includegraphics[width=1.0\columnwidth]{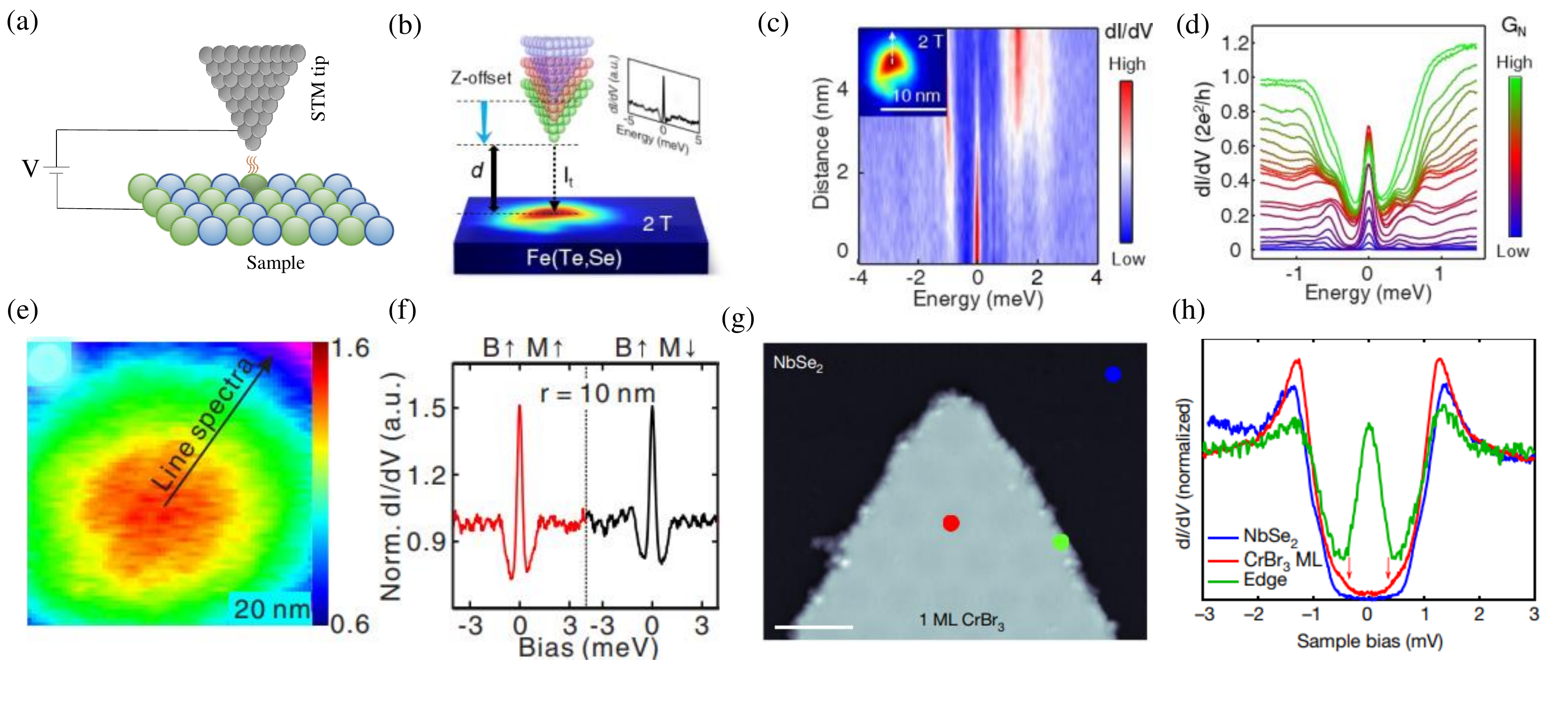}
    \caption{\textbf{Scanning tunneling spectroscopy for TSC studies.} (\textbf{a}) By varying bias voltage, the differential tunneling current becomes a measure of local density states for electrons. (\textbf{b}) A zero-bias conductance map under 2.0 T is shown on a sample surface FeSe$_{0.45}$Te$_{0.55}$. dI/dV spectrum measured at the center of the vortex core. (\textbf{c}) A line-cut intensity plot from the vortex shows a stable MZM across the vortex core. (\textbf{d}) An overlapping plot of dI/dV spectra under different tunnel coupling values. Figures recreated from \cite{zhu2020}. (\textbf{e}) Zero bias mapping of a vortex at 0.1 T with the spin nonpolarized tip on the topological superconductor Bi$_{2}$Te$_{3}$/NbSe$_{2}$. (\textbf{f}) dI/dV away from the center of a vortex measured with a fully spin-polarized tip, where the tunneling is found independent of the spin polarization. Figures recreated from \cite{sun2016}. (\textbf{g}) STM image of a monolayer-thick CrBr$_{3}$ island grown on NbSe$_{2}$. (\textbf{h}) Experimental dI/dV spectroscopy on the NbSe$_{2}$ substrate (blue), the middle of the CrBr$_{3}$ island (red), and at the edge of the CrBr$_{3}$ island (green). Figures recreated from \cite{kezilebieke2020}.}
    \label{Fig4}
\end{figure}

One of the crucial tools for identifying TSCs and exploring their properties is STM/STS. This powerful method is well suited for the study of TSCs because it has high spatial and energy resolution of electron spectra and atomic topography (schematic in Fig. \ref{Fig4}a). One of the main utilities of this technique is in investigating electronic bandstructures through quasiparticle interference (QPI). QPI measures the quasiparticle spectrum by imaging electronic standing waves through differential conductance maps. This method has been used to identify pairing symmetries \cite{hanaguri2010} and topological surface states \cite{zhang2018} in TSC candidates, especially the Fe-based superconducting materials (Fig. \ref{Fig4}b-d). QPI can be advantageous over other bandstructure reconstructing techniques like ARPES because it provides atomic-scale sensitivity to different surfaces and can be operated in a finite magnetic field.  

STM/STS is also prominently used to search for MZMs localized within vortex cores and at other defects such as step edges and chain boundaries. In Fe-based systems, STM has been extensively used to characterize zero-bias peaks at the center of superconducting vortices \cite{wang2018,liu2018,zhu2020,machida2019,kong2019}. Crucial for these investigations is the ability of the STM to identify vortex cores with atomic resolution and to measure their differential conductance within the center of the vortex at a finite magnetic field. In this way, the zero-bias conductance can be measured as a function of distance from the center, enabling a direct comparison to theoretically predicted MBS profiles (for example, Fe(Te, Se) in Fig. \ref{Fig4}b). STM/STS will also likely play a key role in investigating Majorana braiding operations, which are the foundation for topological quantum computing schemes, by manipulating MBSs directly with the tip \cite{wang2018}. 		
 It is also possible to arrange magnetic atoms on the surface of conventional superconductors to directly realize the 1D Kitaev chain model, which prominently features MBSs at each end of the chain \cite{nadj-perge2014,ruby2017,kim2018}. In these experiments, the STM tip writes ferromagnetic atoms into a chain along the surface of a superconductor. Next, it can directly probe the tunneling density of states at either end of the atomic chain and characterize the ZBPs as a function of distance from the chain boundary and tip spin polarization \cite{jeon2017}. 
 
Beyond investigating the possibility of MZMs and TSC in vortex cores  and magnetic chains on superconducting substrates, there have also been STM explorations of 1D topological edge-states in 2D heterostructures (such as Fig. \ref{Fig4}e-h). These heterostructures aim to proximitize superconductivity into topological materials, which have 1D helical edge modes \cite{jack2019,lupke2020}. In one study, higher-order topological insulator Bi was grown on superconducting Nb, and the STM was used to arrange Fe atoms in a chain along the chiral hinge mode, leading to the observation of MZMs within this chain \cite{jack2019}. In another study, van der Waals heterostructures consisting of atomically thin quantum spin Hall insulator WTe$_{2}$ and superconductor NbSe$_{2}$ were prepared, and a superconducting gap was observed along the 1D edge states of the WTe$_{2}$ flakes \cite{lupke2020}. These experiments highlight the utility of STM in probing TSC and MZM’s with atomic resolution in a diverse set of experimentally realizable materials systems. 

\subsection{Photoemission Spectroscopy}
 ARPES has served as an indispensable technique to measure phenomena related to the collective behavior of electrons and their interactions, namely over the course of different eras of superconductivity research. For example, extensive research on iron-based superconductors has widely been enhanced due to the refined orbital information of the electronic states and momentum-resolved electron dynamics provided by ARPES. In that regard, significant improvements to energy and momentum resolution in ARPES in recent decades, made possible with the latest laser and synchrotron-based light sources, have enabled the measurements of quantities such as the superconducting energy gap and bandstructures with unprecedented precision \cite{lv2019,sobota2021}. More recently, precise ARPES measurements of the electronic bandstructure of topological materials have unveiled topological Dirac and Weyl band crossings in the bulk of the materials, in addition to corresponding topological surface states. In light of these topological states, the quest for TSC evidently follows two routes, as mentioned previously, and ARPES has served as key probes in both cases.

\begin{figure}[ht!]
    \centering
    \includegraphics[width=1.0\columnwidth]{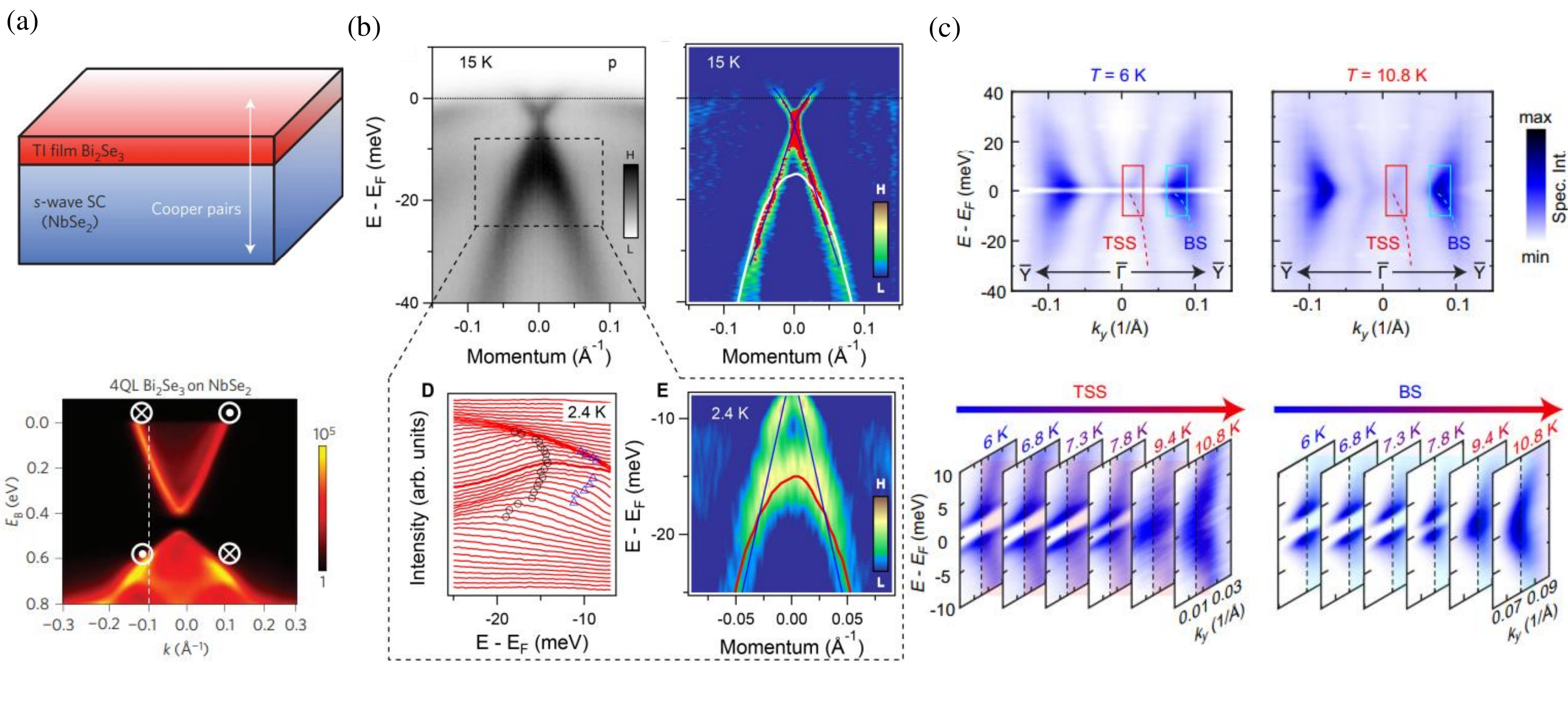}
    \caption{\textbf{ARPES studies on TSC.} (\textbf{a}) A schematic diagram of ultrathin Bi$_{2}$Se$_{3}$ films epitaxially grown on the (001) surface of $s$-wave superconductor 2H–NbSe$_{2}$ (top). High-resolution ARPES dispersion map of Bi$_{2}$Se$_{3}$ film on NbSe$_{2}$ where the white circle and cross schematically show the measured direction of the spin texture on the top surface of Bi$_{2}$Se$_{3}$ film (bottom). Figure recreated from Ref. \cite{xu2014-1}. (\textbf{b}) Band dispersion of FeTe$_{0.5}$Se$_{0.5}$ (top). The momentum distribution curvature plot shows the Dirac-cone type band. The Dirac-cone type band (blue lines) is the topological surface band, and the parabolic band (white curve) is the bulk valence band. In the low temperature (2.4 K) data, the spectral features are narrower. The extracted bands overlap well with the curvature intensity plot, confirming the existence of the parabolic bulk band and the Dirac-cone-type surface band (bottom). Figure recreated from Ref. \cite{zhang2018}. (\textbf{c}) Photoemission spectra intensity plots of the band dispersions in the superconducting (top left) and normal (top right) states show clear superconducting gaps from both the TSS (red dashed lines) and bulk state (BS) (blue dashed lines) in 2M-WS$_{2}$. Temperature dependence of the band dispersions of the TSS and BS show the clear superconducting gap below $T_{c}$ (bottom). Figure recreated from Ref. \cite{li2021}.}
 \label{Fig5}
\end{figure}

One method of choice pioneered during the earlier discoveries of the topological matter is the doping of TIs to induce superconductivity or at the interface of fabricated heterostructures of TIs with superconductors via the proximity effect. In this case, driving the non-spin degenerate surface states of the three-dimensional TI towards superconductivity would induce the realization of a spinless $p_x + ip_y$ model with preservation of TRS, as described by the Fu-Kane model \cite{fu2008}. The earliest measurements of TI Bi$_2$Se$_3$ doped with copper using ARPES revealed spin-polarized topological surface states preserved at the Fermi level in the superconducting regime \cite{wray2010}. This was followed by works on heterostructures of Bi$_2$Se$_3$ thin films on NbSe$_2$ \cite{wang2012, xu2014-1} or on Bi$_2$Sr$_2$CaCu$_2$O$_{8+\delta}$ \cite{wang2013, yilmaz2014}. While the ARPES measurements on the NbSe$_2$ substrate (Fig. \ref{Fig5}a) samples reveal surface state Dirac cones with an appreciable hybridization-related energy gap, there are disagreeing views on whether the proximity effect is suppressed for the samples on the cuprate superconductor substrate due to short coherence lengths, among other reasons. Follow-up studies on Dirac cone surface states in doped Bi$_2$Se$_3$ with Tl \cite{trang2016} or Sr \cite{almoalem2021}, the emergence of isotropic superconducting gaps of samples with Pb(111) grown on similar Tl-doped Bi$_2$Se$_3$ thin films \cite{trang2020} and helical Cooper pairing through measurements of the superconducting gaps in heterostructures of Bi$_2$Se$_3$ on Nb \cite{flototto2018} via ARPES have revealed how this momentum-resolved probe can be used to establish these systems as potential platforms for two-dimensional TSC (Fig. \ref{Fig5}c and d).

Concurrently, there are tremendous efforts to uncover materials that are fully-gapped bulk superconductors and that inherently possess strong nontrivial topological surface states, the amalgamation of which may serve as a potential platform to induce topological superconductivity through the Fu-Kane paradigm. ARPES has undoubtedly participated at the forefront of the discovery of these materials, serving simultaneously as a probe of the induced superconducting gap and of the topological surface states by virtue of modernistic upgrades in energy and momentum resolution. To date, the strongest evidence for topological superconductivity have originated from the observation of topological spin-helical surface Dirac cones on the (001) surface of FeTe$_{0.55}$Se$_{0.45}$ \cite{zhang2018} (Fig. \ref{Fig5}b). Evidenced from calculations of the topological order manifesting the spin-orbit-coupling-induced band inversion from Se substitution and previous work \cite{wang2015}, ARPES measurements reveal the superconductivity induced in the topological surface states as the system enters the superconducting state through superconducting gaps of 1.8 meV which is isotropic in momentum. Beyond FeTe$_{0.55}$Se$_{0.45}$, there are other candidate systems such as binary Pd-Bi systems \cite{sakano2015, neupane2016}, TaSe$_3$ \cite{chen2020}, 2M-WS$_2$ (Fig. \ref{Fig5}c) \cite{li2021}, Li(Fe, Co)As \cite{zhang2019}, and CaKFe$_4$As$_4$ \cite{liu2020} which have been measured for the topological surface states, yet typically suffer from overlapping bulk bands with small gaps which are difficult to resolve in ARPES data in comparison to the Bi$_2$Se$_3$-based systems. They remain to be validated with other probes to establish their topological states and determine whether they can serve as platforms for realizing Majorana zero modes. 

\subsection{Transport Measurements}

The chiral Majorana edge modes in a $p+ip$ TSC is expected to provide direct thermal transport evidence on the presence of Majorana fermions. However, such measurement has only been done in quantum spin liquid candidate, a system with long-range entanglement (Fig. \ref{Fig1_TSC_Class}a) but can also host Majorana edge modes \cite{Yokoi_2021_QSL_thermal}. Even so, other electrical transport measurements could still provide some insights. Some of the transport measurements in natural TSC candidates include Shubnikov-de Haas oscillations (SdHOs) that show a nontrivial Berry phase shift. In T$_d$-MoTe$_2$, the Landau level index plot shows a $\pi$ Berry phase shift \cite{luo2016}. In another candidate, Sr$_x$Bi$_2$Se$_3$ SdHOs confirm the shift expected from a Dirac spectrum, giving transport evidence of surface states \cite{liu2015}. Magnetotransport measurements show striking SdHOs in other putative TSC candidate TaSe$_3$ \cite{xia2020}, LuPdBi \cite{pavlosiuk2015}, and YPtBi \cite{pavlosiuk2016}. According to Abrikosov's theory of quantum magnetoresistance, linear magnetoresistance in zero-gap band systems with a linear energy dispersion is a result of the system being in the extreme quantum limit thereby confining all the electrons in lowest Landau level \cite{pavlosiuk2015, abrikosov1998}. In both LuPdBi and YPtBi, the measured zero field resistivities were fitted with a sum of theoretical metallic surface states and semiconducting bulk components and found to agree very well, providing the signature of the existence of nontrivial surface states  ~\cite{pavlosiuk2015,pavlosiuk2016}. The thermal conductivity measurements conducted at low temperatures indicate that 2M-WS$_2$ may possess either an anisotropic superconducting gap or multiple nodeless superconducting gaps, consistent with features of TSC candidates \cite{wang2020}.

In addition to bulk transport, an alternative local probe compared to STS for probing TSC is quantum point contact spectroscopy (PCS). Similar to STS, PCS can be employed to acquire the transport signature of MZMs in ultra thin interfaces~\cite{lee2019}, as well as bulk topological crystalline interfaces such as Dirac semimetal Cd$_3$As$_2$~\cite{wang2016-1}, Weyl semimetal TaAs~\cite{wang2017} and TSC candidates Sr$_2$RuO$_4$~\cite{wang2015-2} and Au$_2$Pb~\cite{xing2016}. In single crystalline samples, a needle anvil PCS configuration is found to be convenient. In PCS, two major types of induced superconductivity are observed in the interfaces. Tip-enhanced superconductivity (TESC) is observed when a superconductor is in contact with a normal metal tip at the interface, whereas tip-induced superconductivity (TISC) is observed when nonsuperconducting tip contacts with a nonsuperconducting material in the interface \cite{wang2018-2}. PCS could also be employed to study multiband superconductivity \cite{daghero2010}.

Another alternative transport characterization technique for the TSC is the Josephson junction. Josephson junctions can be used to probe the phase coherence of a superconductor. By measuring the Josephson current as a function of the applied voltage or magnetic field, one can obtain information about the superconducting properties of the material \cite{schuray2017,schuray2020}. The presence of a MBS may lead to a 4$\pi$-periodic supercurrent through a Josephson junction. A systematic study of the radio frequency response for various temperatures and frequencies conducted by de Ronde \textit{et al.} has resulted in the observation of a 4$\pi$-periodic contribution to the supercurrent in Josephson junctions based on BiSbTeSe$_2$ \cite{nano10040794}. Such measurements can provide evidence for the existence of topological properties in the material, even if they do not directly confirm topological superconductivity. 

\subsection{Muon Spin Spectroscopy}

\begin{figure}[ht!]
    \centering
    \includegraphics[width=1.0\columnwidth]{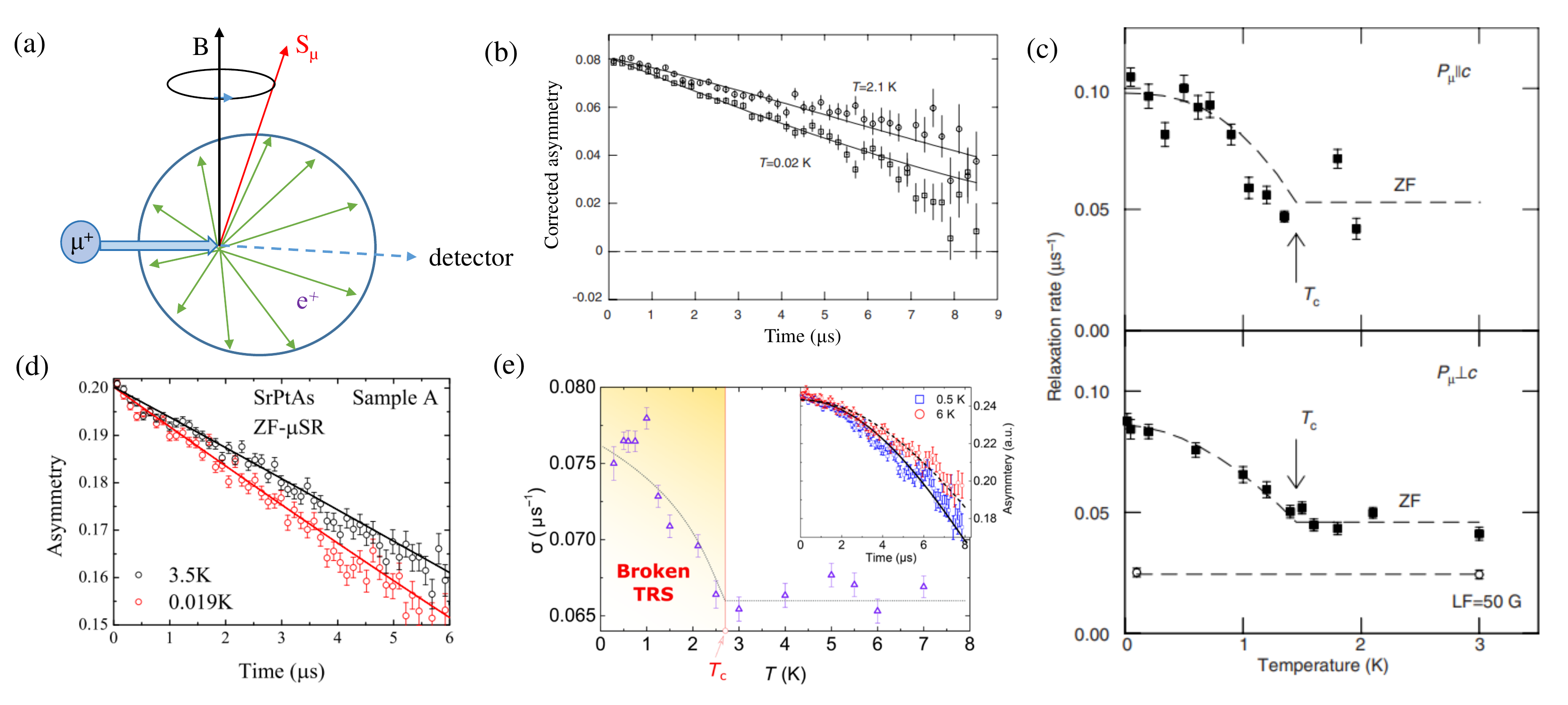}
    \caption{\textbf{Signature of TSC from $\mu$SR.} (\textbf{a}) The symmetric diagram of positron emission and the muon spin direction. (\textbf{b}) Time evolution of the spin polarization of muons above and below superconducting transition temperature under zero-field (ZF) conditions indicates the TRS breaking for Sr$_{2}$RuO$_{4}$. (\textbf{c}) ZF muon relaxation rate for the initial muon spin polarization for Sr$_{2}$RuO$_{4}$. Figures recreated from Refs. \cite{luke1998}. Time evolution of the muon spin polarization in ZF conditions suggests broken TRS for (\textbf{d}) SrPtAs (Figure recreated from Ref. \cite{biswas2013}) and (\textbf{e}) 4Hb-TaS$_{2}$ (Figure recreated from Ref. \cite{ribak2020}).}    
    \label{Fig6}
\end{figure}

Muon spin spectroscopy ($\mu$SR) is an extremely sensitive local probe to microscopically resolve the pairing symmetry in superconductors \cite{lee1999,nagamine2003,blundell1999}. In this experimental technique, 100\% spin-polarized positive muons are implanted in the material and are used to detect the corresponding muon spin evolution with time on an atomic-scale limit (Fig. \ref{Fig6}a). The precession of muon spin is due to the magnetic field due to its local environment, similar to other magnetic resonance techniques, such as nuclear magnetic resonance \cite{hore1995} and electron spin resonance \cite{wertz1986}. In the mixed or vortex state of type-II superconductors, a $\mu$SR study gives rise to a spatial distribution of local magnetic fields, which demonstrate itself through a relaxation of the muon polarization \cite{lee1999,nagamine2003}. Transverse field-$\mu$SR measurements reveal one of the most important superconducting parameters, London penetration depth, that is inversely related to the density of Cooper pairs n$_{s}$ \cite{dunlap2019}. The temperature and
the magnetic field variation of n$_{s}$ directly indicate the symmetry of the superconducting gap. In the zero-field (ZF) configuration, $\mu$SR is a wonderful tool to reveal the broken TRS in the superconducting state that has significant consequences for the symmetry of pairing and quasi-particle spectrum \cite{lee1999, nagamine2003, blundell1999}. Because TSC are classified with respect to the TRS symmetry, such as TRS preserved TSC (or helical superconductor) and TRS broken TSC (or chiral superconductor) \cite{kallin2016}, $\mu$SR is frequently used to identify TSC candidates.

In non-centrosymmetric superconductors (such as CePt$_{3}$Si \cite{bauer2004}, Li$_{2}$(Pd$_{1-x}$Pt$_{x}$)$_{3}$B \cite{badica2005}), broken inversion symmetry allows for Rashba and Dresselhaus SOC that lift the spin degeneracy and split the Fermi surface. In such a system, parity is ill-defined, and hence mixing of spin-singlet ($s$-wave) and spin-triplet ($p$-wave) states is allowed \cite{sato2009,tanaka2009}. If the $p$-wave gap is larger than the $s$-wave gap in a 2D non-centrosymmetric superconductor, then topological properties may appear \cite{sato2009,tanaka2009}. In such a topological state, if it preserves TRS, helical Majorana fermions show up at the edge \cite{badica2005}.

In a chiral superconductor, the phase of the complex superconducting gap function winds in a clockwise or anti-clockwise sense as momentum vector, $\Vec{k}$ moves about some axis on the Fermi surface \cite{kallin2016}. The gap function breaks TRS spontaneously and is degenerate with its time-reversed partner. It is a type of topological state and carries certain signatures of its non-trivial topology. The vortex core of a chiral $p$-wave superconductor exhibits a single MZM for the case of spinless fermions \cite{kallin2016}. Many materials such as Sr$_{2}$RuO$_{4}$ (full gap, chiral-$p$) (Fig. \ref{Fig6}b and c) \cite{luke1998}, SrPtAs (full gap, chiral-$d$) (Fig. \ref{Fig6}d) \cite{fischer2014, biswas2013}, 4Hb-TaS$_{2}$ (Fig. \ref{Fig6}e) \cite{ribak2020}, UTe$_{2}$ (chiral-$p$) \cite{ran2019,ran2019-2}, UPt$_{3}$ (nodal gap, chiral-$f$) \cite{avers2020}, and LaPt$_{3}$P (chiral-$s$) \cite{biswas2021} are predicted as chiral superconductors, but are still in debate due to the lack of direct evidence of a MZM. The heavy fermion, URu$_{2}$Si$_{2}$ \cite{li2013,okazaki2011}, displays a mysterious ``hidden order" phase with broken TRS, indicating a possible chiral $d$-wave state. Other interesting systems such as water-doped cobaltate, Na$_{x}$CoO$_{2}$$\cdot$yH$_{2}$O, (x = 0.3, y = 1.3) \cite{takada2003}, twisted double-layer copper oxides \cite{can2021}, and doped graphene \cite{nandkishore2012} are proposed as chiral superconductors, where $\mu$SR and polar Kerr experiments would be great interest to look for possible broken TRS.

\section{Future Prospective}\label{sec13}


Despite a decade of extensive searching, only a few TSC candidates have been identified (Table \ref{tab1}). Given the great promise of MZM-based topological quantum computation in a TSC, there is an urgent need for fundamentally new approaches to accelerate the search and identification of new TSC materials.

On the one hand, the experimental identification of TSC is challenging due to the elusive experimental fingerprints it presents. For instance, both MBS and spurious signals such as Andreev bound states can create the zero-biased peak measured with tunneling spectroscopy. One approach to address this issue is to improve the current experiments by enlarging the measurement parameter space. Recently, Ziesen et al. proposed a new approach that involves replacing single-shot STS with a sequence of shots at varied system parameters, which has the potential to significantly improve the identification of MZM \cite{Statistical_MBS_PRL2023}. This strategy could have general implications, as it involves redesigning the probe without the need for drastic changes to the existing experimental apparatus. Beyond improving existing experimental configurations, completely new experimental configurations may be needed in the future to provide more conclusive evidence and efficiently and reliably screen out TSC candidates. 

On the other hand, modern computational methods have revolutionized many branches of materials science and have accelerated the search for new materials. However, computational-aided TSC search is challenging. The popular Density Functional Theory (DFT) calculations, which have been highly successful in searching for materials with bandstructures and topology \cite{Regnault2022,Vergniory2022}, are single-particle in nature and rely on local or extended single-particle basis sets. As a result, they are not inherently designed to account for paired states in superconductors. Recently, the use of symmetry indicators has shown great promise in identifying band topology \cite{Tang2019a,Tang2019b}, and corresponding symmetry indicators for TSC have also been developed \cite{ono2019symmetry}. However, obtaining the pairing symmetry - the crucial input needed to apply the symmetry indicator for TSC - is not readily accessible by any means, which impedes the use of symmetry indicators to search for TSC candidates computationally.

Machine learning (ML) has emerged as a powerful tool for materials discovery, including the search for quantum materials \cite{stanev2021artificial}. However, due to the data-driven nature of ML and the absence of confirmed TSC materials \cite{li2021} as well as reliable simulation methods, popular ML methods such as classification, clustering, and generative models may not be suitable for searching for TSC, mainly due to the out-of-distribution (OOD) problem \cite{Bulusu2020AnomalousED}. Nevertheless, ML can still be useful in certain scenarios. Recent studies have shown that ML can optimize Majorana wire gate arrays towards improved topological signatures with reduced disorder effects \cite{Thamm_2023_PRL}.

Despite the OOD problem, we can leverage various OOD detection models through ensemble learning \cite{DEANGELI2022103957} as a confidence predictor for absent, null phases. This predictor can be used to screen the existing material database or incorporated into a generative adversarial model, which has been highly successful in the field of image generation, to generate new candidates for materials not present in the training phases \cite{AGGARWAL2021100004, 10.1093/nsr/nwac111, doi:10.1021/acsomega.2c03264}. In the longer term, as TSC materials need to be confirmed, popular ML methods can further accelerate predictions for even more TSC candidates, forming a positive feedback loop toward accelerated discovery.

Last but not least, integrating ML with experiments offers an alternative pathway to augment the capability of experimental techniques for throughput and accuracy \cite{Nina2022a,Chen2021}. For instance, it has been shown that the experimental spatial resolution to identify proximity effect - one key approach to realize TSC - can be enhanced by a factor of two through ML-based analysis \cite{Nina2022b}. By increasing the throughput and resolution of experiments in conjunction with ML techniques, the TSC search can also be expected to accelerate in the near future.\\

\backmatter

\bmhead{Acknowledgments}
 MM and NCD acknowledge the support from U.S. Department of Energy (DOE), Office of Science (SC), Basic Energy Sciences (BES) Award No. DE-SC0020148. RO and AC acknowledge support from DOE BES Award No. DE-SC0021940. TN and TB acknowledge National Science Foundation (NSF) Designing Materials to Revolutionize and Engineer our Future (DMREF) Program with Award No. DMR-2118448. TB and ML are partially supported by NSF Convergence Accelerator Award No. 2235945. ML acknowledges the support from Class of 1947 Career Development Professor Chair and support from R Wachnik. 

\bmhead{Declarations}
The authors declare no competing interest.

\newpage
\bibliography{sn-bibliography}

\end{document}